\documentclass[9pt]{book}

\usepackage[dvips]{graphicx,color}
\usepackage{makeidx,universe}
\usepackage{wrapfig}

\makeauthorindex

\BookTitle{The proceedings of the Physics of Accreting Compact Binaries}

\CopyRight{\copyright 2010 by Universal Academy Press, Inc.}

\begin{document} 

\pagenumbering{arabic}

\chapter{%
SDSS J0018+3454:  An Extremely Long Orbital Period CV}

\author{\raggedright \baselineskip=10pt%
{\bf D. Gonzalez, 
G. Tovmassian, 
S. Zharikov, 
A. Aviles 
and 
J. Echevarria} \\ 
{\small \it %
Institute of Astronomy, UNAM, AP 877, Ensenada, Baja California, 22800 Mexico
}
}


\AuthorContents{Daisaku Nogami, Paul Mason, and Christian Knigge} 

\AuthorIndex{Nogami}{D.} 

\AuthorIndex{Mason}{P.A.} 

\AuthorIndex{Knigge}{C.} 

     \baselineskip=10pt
     \parindent=10pt

\section*{Abstract} 

We present here results of an optical spectroscopic study of a new Cataclysmic Variable SDSS J001856.93+345444.3.  
We demonstrate that the most probable value of the orbital period of the system is P$_{\mathrm {orb}}=0.6051\pm0.022$ days (=14.5226 hours), based on the measurements of radial velocity of a complex of absorption features  emanating from the K2-K4\,V type secondary component. However, the radial velocity measurements from the emission lines are best folded with the period P$_{\mathrm {em}} = 0.5743$\,day (=13.78 hours). The $\gamma$-velocity of the emission lines varies significantly from epoch to epoch. There is  an underlying broader and weaker component to the emission lines, which we could not resolve.  Based on the appearance of the emission lines, the presence of very strong He {\sc ii} lines and the moderate polarization detected by Dillon et al. (2008), we conclude that SDSS J0018+3454 is an asynchronous magnetic CV (Polar).

\section{Introduction} 

\begin{figure}[t]
  \begin{center}
  \includegraphics[height=13pc]{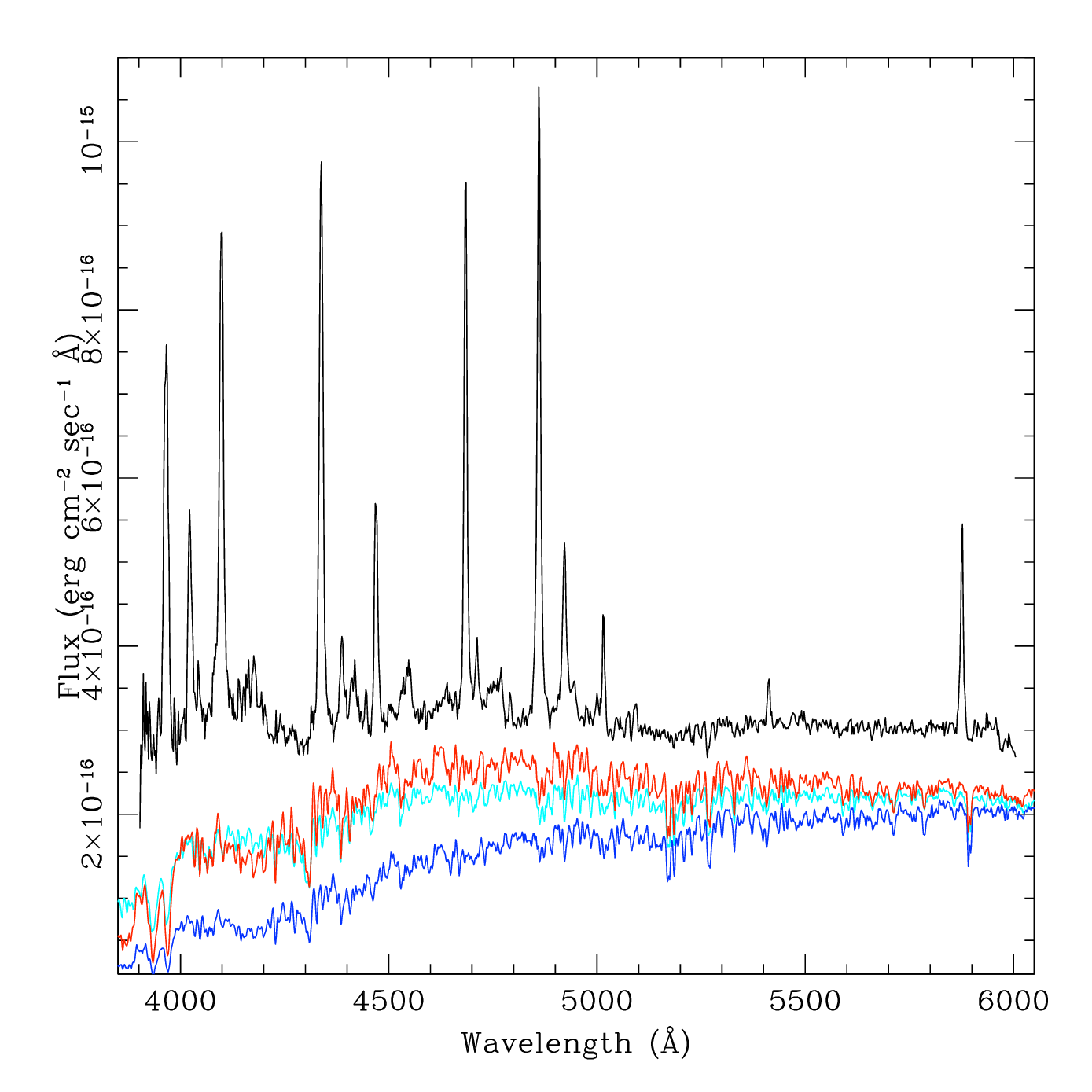}
  \end{center}
  \vspace{-1pc}
  \caption{The combined spectrum of SDSS0018+3454, after individual spectra were corrected for radial velocity variations (derived from the absorption line measurements  by cross correlation). 
For comparison  main-sequence star spectra are over-plotted:
G9\,{\sc v} - red line,
KO\,{\sc v} - cyan line,
K5\,{\sc v} - blue line
}
\end{figure}

\begin{table}[t]
\small
    \caption{Log  of Observations.} 
\begin{center}
\begin{tabular}{l|ccccc} \hline
Date  & Exposure  &   Resolution & Range &   No of spectra       & Total      \\ 
           &     sec         &        FWHM   \AA                         &       \AA          &                                   &   hours \\    \hline
06/11/08    & 1200    & 4.1 & 3900-5950 &  22 &   8        \\
07/11/08     & 1200 & 4.1   & 3900-5950 &  19     & 7      \\
08/11/08     & 1200 & 4.1   & 3900-5950  & 25     & 9      \\
09/11/08     & 1200 & 4.1  & 3900-5950   &  10     & 4       \\ 
11/11/08    &  1200  &  4.1 &  6050-8100 &  10  &  4 \\
06/12/08        & 1200 &  4.1 & 3825-5875    & 22 &   8        \\
07/12/08     & 1200 & 4.1 & 3850-5900 & 9     & 3      \\
28/08/09     & 1200 & 4.1 & 3825-5875 & 17     & 7      \\
29/08/09     & 1200 & 4.1 & 3825-5875 & 10     & 4       \\  
29/09/09     & 1200 & 4.1 & 3850-5900 &  22     & 8       \\ 
\hline
\end{tabular}
\end{center}
\vspace{-1pc}

\end{table}

SDSS J001856.93+345444.3 was identified as a Cataclysmic Variable  by Szkody et al., (2005).  The object has narrow Balmer emission lines, He\,{\sc i} and  strong He\,{\sc ii}.  
They took two hour follow up observations and could not detect any significant  radial velocity variations that would yield the orbital period.
Similarly, Dillon et al., (2008), could not find any photometric variability with times corresponding to the orbital period in their four nights of observations. But they point out the presence of absorption features and conclude that the system must be a long period CV with low inclination. Szkody et al., (2005) also performed polarimetry and found no significant polarization within those two hours.

\section{Observations} 

Observations of SDSS J0018+3454 were performed on the 2.1 m telescope at the  Observatorio Astronomico Nacional San Pedro Martir, in Baja California Mexico.  The Boller \& Chivens spectrograph was deployed with the 600 l/mm grating element, coupled with the $1024\times1024$\,pix SITe CCD, producing a FWHM resolution of 4.1\AA.  The wavelength calibration was obtained using  arc lamp taken after every tenth exposure and spectrograph flexing with azimuth was accounted for  by correcting the reference point of each spectrum  using night sky lines. Most of the spectra were taken in the 3900-6000\,\AA\ range, except the spectra taken on 6 Dec 2008, which covered H$_\alpha$ and redward. The latter spectra were not flux calibrated. The rest of the data were calibrated using spectrophotometric standard stars observed on the same night.  The standard long slit reduction of the data was made using IRAF procedures.

\begin{figure}[t]
  \begin{center}
  \includegraphics[height=13pc]{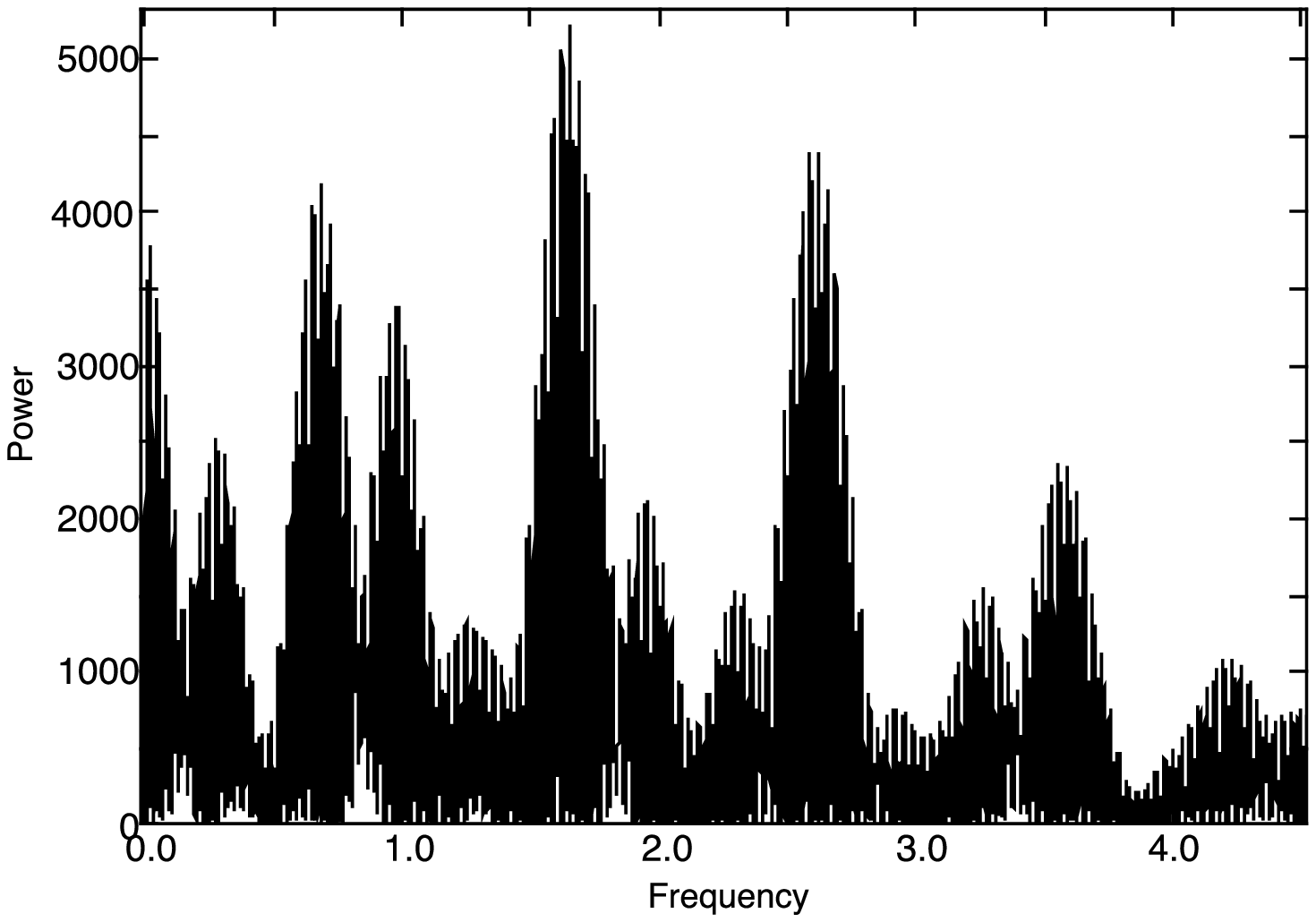}
    \includegraphics[height=16pc]{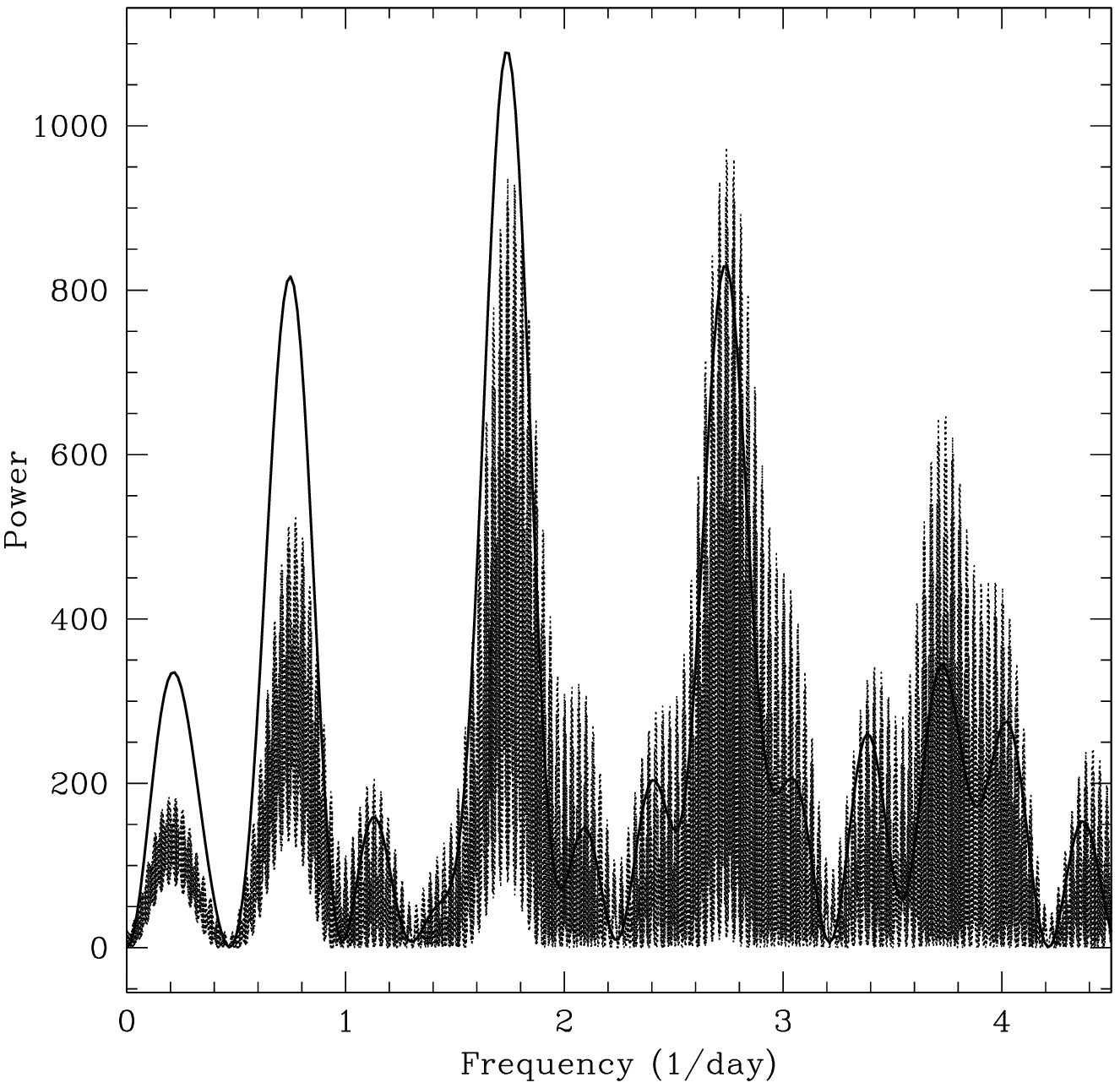}
  \end{center}
  \vspace{-2pc}
  \caption{ The power spectra of the RV measurements. Top panel:  for the  absorption lines.
  The strongest peak is at 1.65 day$^{-1}$.
Bottom:  for the emission lines. The dominant peak is at 1.74 day$^{-1}$.
The solid line is based on data from only 2008.
}
\end{figure}

\section{Results} 

The spectra of SDSS J0018+3454 (Fig 1) we obtained was over a much longer time span than  was done previously, confirming that the emission lines are narrow and single peaked, that He\,{\sc ii} is exceptionally strong and that there is a significant presence of secondary star absorption features in the spectra. We used reference spectra from Jacoby et al., (1984) to measure radial velocities of absorption features in the wavelength range 5100-5850 \AA\  and by excluding the He\,{\sc ii} 5411\AA\ from the analysis. Three different spectral classification standards (K0, K4 \& K5\,{\sc v}) were used. The XCSAO procedure from the RVSAO package (Kurtz \& Mink, 1998) was used for the measurements. They all produced similar results, but the K4\,{\sc v} spectrum converged with most of the observed spectra. Apparently, at some phases, when the secondary star faced the observer, the cross-correlation procedure was not able to pick up the correct solution. The period analysis of the radial velocities by means of the  DFT produced the strongest peak at 1.6526 day$^{-1}$ frequency (Fig 2). The corresponding orbital period of 14.52 hours is one of the longest known. The radial velocity curve folded with this period is presented in the upper panel of Fig 3.  The best fit of RV = $\gamma + K_{\mathrm {abs}}  sin (2\pi/P_{\mathrm {orb}}$) occurs when K$_{\mathrm {abs}} =  68$ km/sec. The $\gamma$ value is not really relevant, because is measured relative to a random star. 

Interestingly, radial velocity measurements of  emission lines do not correlate with the absorption. The RV of H$_\beta$ folded with the period derived from the absorption features does not show any trend (bottom panel of Fig 3).  It appears that the velocities measured in 2009 deviate from the 2008 data by about 100 km/sec. The wavelength calibration can not be an issue here,  because the absorption lines do not show any shifts. We note that there is a broad and less intense emission component of the lines, which  is impossible to separate at the given spectral resolution.  What is interesting however, is that after being normalized  to the same $\gamma$-velocity, the RV analysis  of the emission lines shows a peak in the power spectrum at 13.8 hours, which is  5\% shorter than the period obtained from the analysis of the absorption lines. The corresponding power spectrum and folded RV curve are shown on Fig 4 \& 5.

\begin{figure}[t]
  \begin{center}
  \includegraphics[height=14pc]{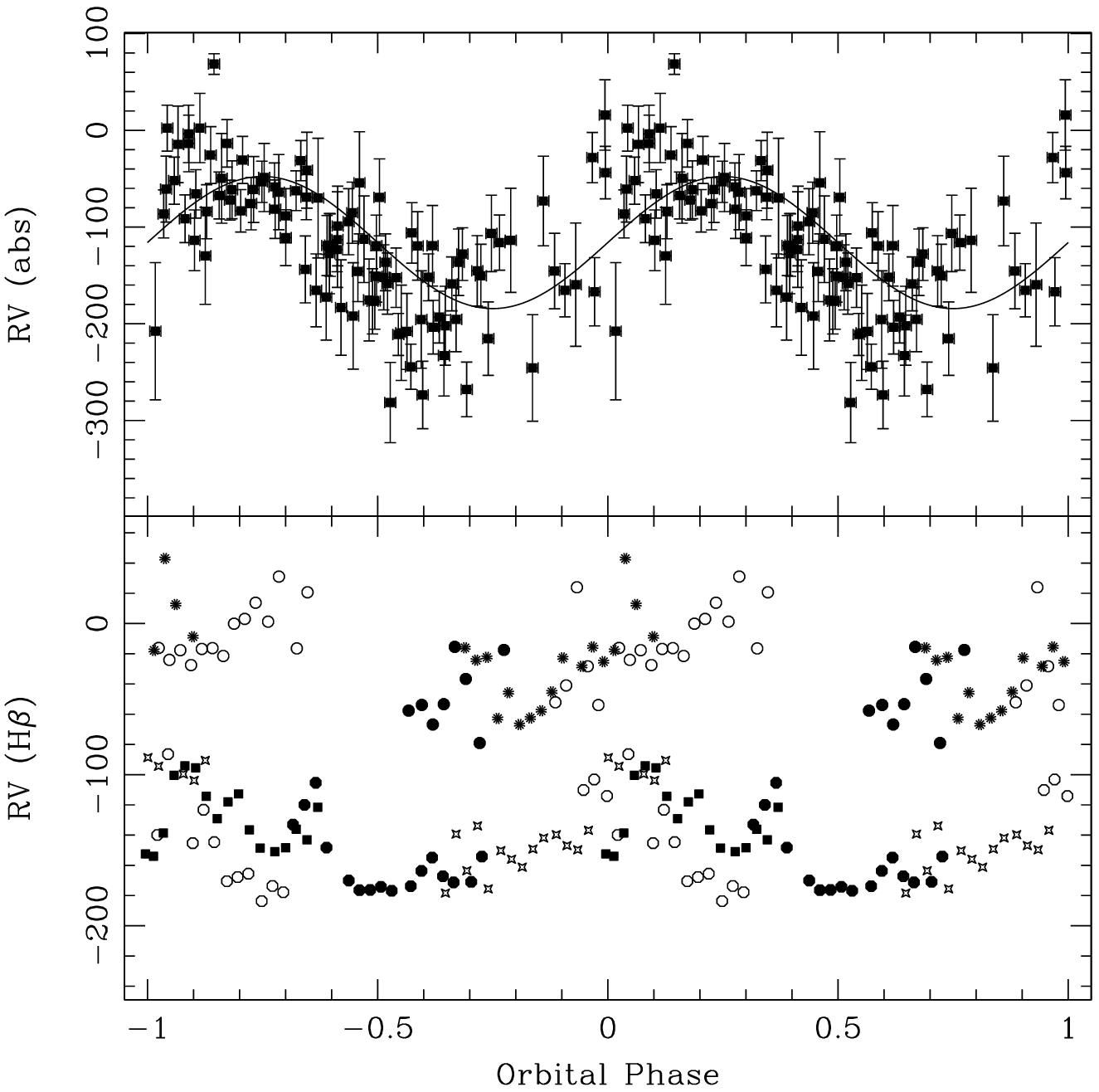}
    \includegraphics[height=14pc]{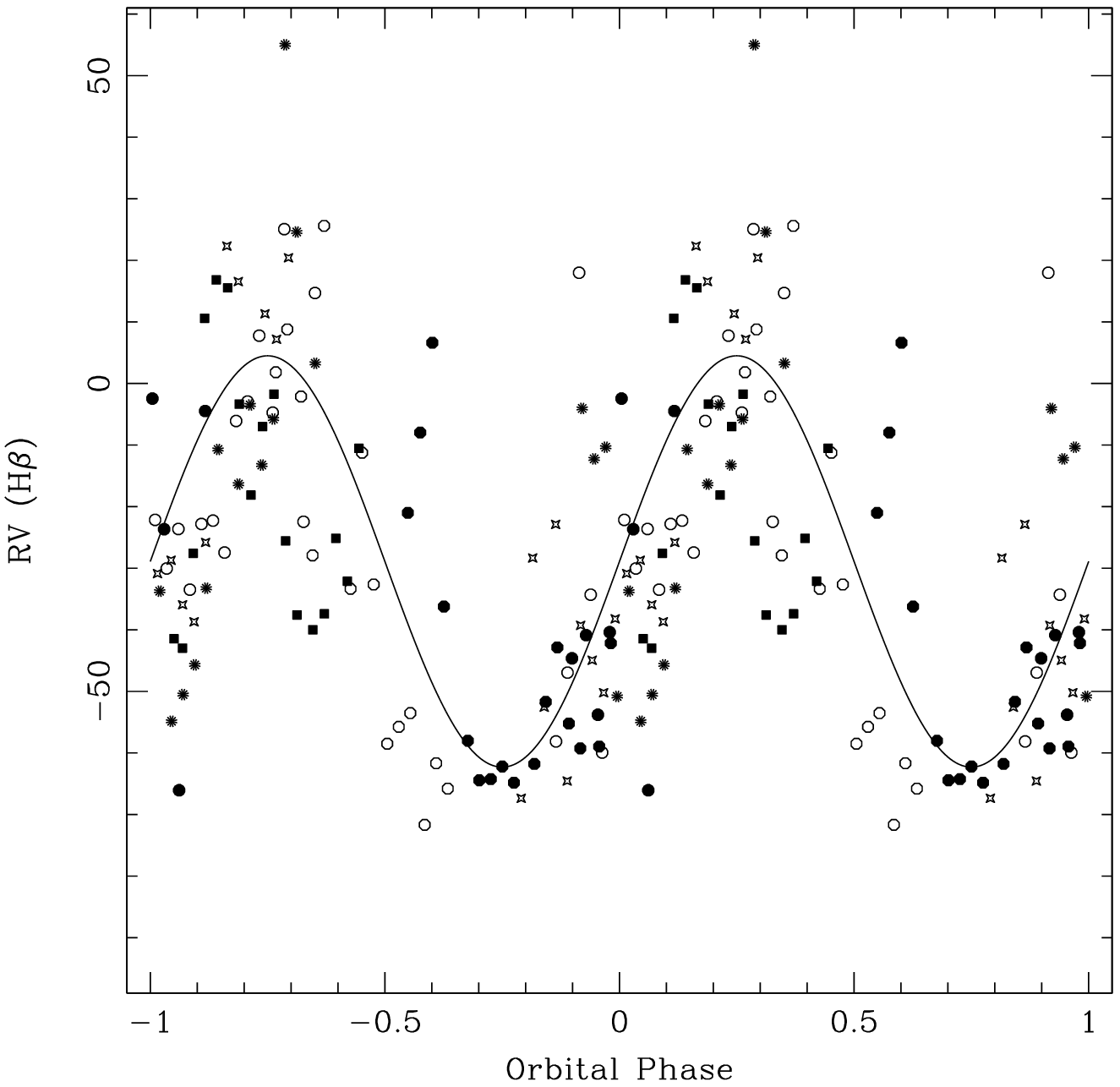}
  \end{center}
  \vspace{-1pc}
  \caption{ left panel: The RV curve folded with the 14.5 hour period, believed to be the orbital perod. Upper panel are absorption features, at the bottom is H$_\beta$ emission.
 right panel: The RV curve of the H$_\beta$ line folded with the 13.8 hour period. The data from 2008 were shifted by ~ 100 km/sec before the period analysis and folding was performed.
}
\end{figure}

\section{Conclusions} 

From the appearance and behavior of the emission lines we conclude that they originate in a mass transfer stream rather than in the accretion disk,  and thus, have a large radial velocity amplitude and are variable in time, as often observed in Polars. The absorption features  are a better indicator of the motion of the secondary star and the orbital period of the system. SDDS J0018+3454 joins a few other similar systems that have  periods $> 12$\,hours, which is characteristic of magnetic CVs. This particular system may has not reached full synchronization yet and needs further study.

\section{References} 

\end{document}